\documentclass[aps,pra,twocolumn, 10pt, showpacs,superscriptaddress,groupedaddress]{revtex4-2} 

\usepackage{subcaption}
\usepackage{braket}
\usepackage[abs]{overpic}
\usepackage{graphicx}
\usepackage{epstopdf}
\usepackage{setspace}
\usepackage{bbold}
\usepackage{comment}
\usepackage{color}
\usepackage{soul}
\usepackage[abs]{overpic}
\usepackage{amsmath}
\usepackage{lipsum}
\usepackage[abs]{overpic}
\usepackage{amsmath}
\usepackage{nicefrac}

\begin{document}

%\preprint{APS/123-QED}

\title{Observation of nonlinear spin dynamics and squeezing in a BEC using dynamic decoupling}

\author{Hagai Edri, Boaz Raz, Gavriel Fleurov, Roee Ozeri and Nir Davidson}
\affiliation{ Department of Physics of Complex Systems, Weizmann Institute of Science, Rehovot 7610001, Israel}

%\date{\today}

\begin{abstract}
We study the evolution of a Bose-Einstein Condensate (BEC) in a two-state superposition due to inter-state interactions. Using a population imbalanced dynamic decoupling scheme, we measure inter-state interactions while canceling intra-state density shifts and external noise sources. Our measurements show low statistical uncertainties for both magnetic sensitive and insensitive superpositions, indicating that we successfully decoupled our system from strong magnetic noises. We experimentally show that the Bloch sphere representing general superposition states is "twisted" by inter-state interactions, as predicted in  \cite{kitagawa1993squeezed,harber2002effect} and the twist rate depends on the difference between inter-state and intra-state scattering lengths  $a_{22}+a_{11}-2a_{12}$. We use the non-linear spin dynamics to demonstrate squeezing of gaussian noise, showing $2.79 \pm 0.43$ dB squeezing when starting with a noisy state and applying 160 echo pulses, which can be used to increase sensitivity when there are errors in state preparation. Our results allow for a better understanding of inter-atomic potentials in \textsuperscript{87}Rb. Our scheme can be used for spin-squeezing beyond the standard quantum limit and observing polaron physics close to Feshbach resonances, where interactions diverge, and strong magnetic noises are ever present.
\end{abstract}

%\pacs{Valid PACS appear here}
% PACS, the Physics and Astronomy         % Classification Scheme.
%\keywords{Suggested keywords}%Use showkeys class option if keyword         %display desired

\maketitle
%introduction
The state of a two-level system can be represented by a vector on the Bloch sphere. Linear operations on this state can be represented by rotations and contractions of the sphere, generated by unitary and non-unitary operations respectively. In cases where the Bloch sphere represents the average state of an ensemble of two-level systems, interactions can introduce non-linear evolution, represented by torsion and one-axis-twist of the sphere. Non-linear spin dynamics can be used for spin-squeezing and generation of non-classical states \cite{kitagawa1993squeezed}. 

In the cold collision regime, interactions are parameterized by \textit{s}-wave scattering length $a_{ij}$, where $\ket{i},\ket{j}$ are two internal states of the interacting atoms (i.e. Zeeman states or hyperfine levels). Our knowledge  of scattering lengths comes from spectroscopic measurements \cite{harber2002effect,fertig2000measurement}, observations of collective oscillations \cite{egorov2013measurement}, position of Feshbach resonances \cite{PhysRevA.83.042704,PhysRevA.73.040702,kerman2001determination} and thermalization experiments \cite{PhysRevA.76.052704,PhysRevLett.89.053202}, which are used for calibrations of detailed calculations of inter-atomic potentials \cite{van2002interisotope,verhaar2009predicting}. Precise knowledge of inter-state interaction parameters, as well as differences in scattering lengths between different internal states, is important for spin-squeezing \cite{gross2010nonlinear,riedel2010atom}, polaron physics \cite{2012GrimmPolaron,FermiPolaronZwierlein,hu2016bosepolaron,jorgensen2016observationbosepolaron}, BEC solitons \cite{PhysRevLett.111.264101,PhysRevLett.120.063202}, and magnetic dipole-dipole interactions \cite{zou2020magnetic}.

Frequency shifts in population-balanced Ramsey spectroscopy due to mean-field interactions were used to measure intra-state scattering length differences $a_{22}-a_{11}$ for both thermal ultracold bosons \cite{PhysRevLett.70.1771} and BEC \cite{harber2002effect} (their absence was observed for both thermal \cite{PhysRevLett.91.250404} and quantum degenerate \cite{gupta2003radio} ultracold Fermi gases). In population-imbalanced spectroscopy additional frequency shift proportional to $a_{22}+a_{11}-2a_{12}$ and to density differences between states was predicted, but was too small to be measured \cite{harber2002effect}. These measurements were limited to magnetic insensitive transitions, where magnetic noises are largely suppressed. Dynamic decoupling (DD) \cite{lidar2014review} could decouple the system from these magnetic noises and preserve its coherence over long times by applying a set of spin rotations. It also decouples it from intra-state interactions and can not be applied to measure them. 

Dynamic decoupling was demonstrated in NMR \cite{haar1962fluctuation,haeberlen1976high}, spins in solids \cite{du2009preserving,de2010universal,bar2013solid}, ultracold atoms \cite{sagi2010process,almog2011direct}, and trapped ions \cite{kotler2013nonlinear,shaniv2016atomic,biercuk2009optimized}. DD aims to decouple a system from its environment. To measure a signal, one needs to modulate the system itself synchronously with the DD scheme \cite{kotler2011single,shaniv2017quantum,PhysRevLett.106.080802}, in a similar fashion to how a lock-in amplifier operates. However, this modulation can generate noises that are in phase with the DD scheme, and requires control of the signal of interest. 

\begin{figure}[t]
% \centering
        \begin{overpic}
            [width=0.92\linewidth]{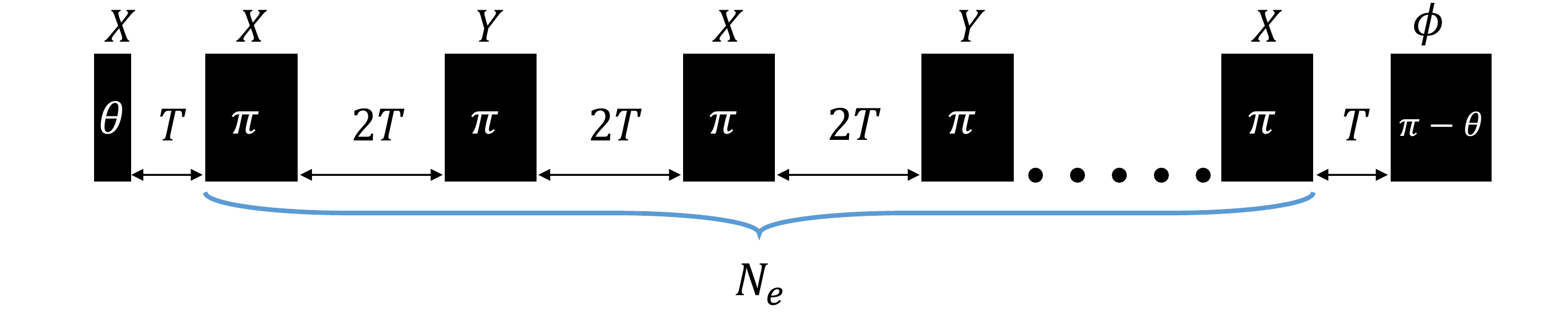}
            \put(12,0){\large \textbf{(a)}}
        \end{overpic}
        \begin{overpic}
            [width=0.92\linewidth]{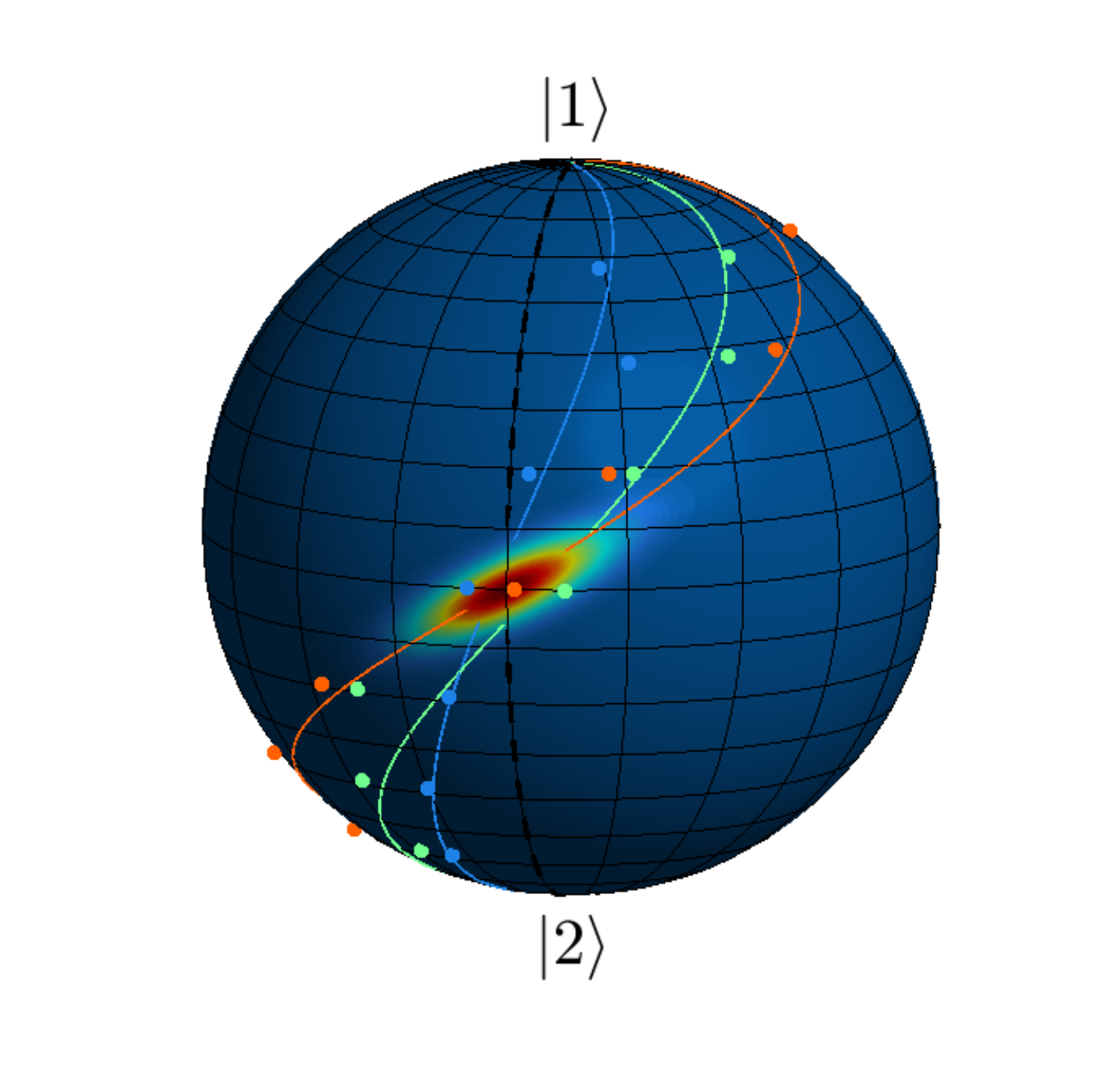}
            \put(12,24){\large \textbf{(b)}}
        \end{overpic}
%    \begin{subfigure}{.45\textwidth}
%        \includegraphics[width=1\textwidth]{pulse_scheme.pdf}
 %      \caption{}
%    \end{subfigure}
%    \begin{subfigure}{.45\textwidth}
%        \includegraphics[width=0.95\textwidth]{twisted_bloch_sphere_final.pdf}
%        \caption{}
%    \end{subfigure}
\caption{\label{fig1} Imbalanced dynamic decoupling. (a) Pulse sequence. An initial pulse of area $\theta$, followed by $N_{e}$ echo pulses around X,Y axes, separated by time $2T$, and a final pulse with area $\pi-\theta$ and phase $\phi$. Initial pulse creates a superposition $\cos{\frac{\theta}{2}}\ket{1}+\sin{\frac{\theta}{2}}\ket{2}$, with density difference $\delta n = n \cos{\theta}$, where $n$ is the total density. (b) A twisted Bloch sphere after imbalanced DD. Measured phase shifts (circles) on magnetic sensitive transition $\ket{1,1}\leftrightarrow\ket{2,0}$ (similar to Fig. \ref{fig4}) at several latitudes for increasing total evolution times (blue - 12.1 ms, green - 26.2 ms, orange - 54.6 ms). Solid lines are fits to the data. Locally, a coherent state around the equator is squeezed by one-axis-twist, illustrated by the colored distribution.}
\end{figure}

In this letter, we propose and demonstrate a method for measuring interactions in ultracold gases in a noisy environment with a long coherence time and increased sensitivity to inter-state interactions. Our measurements are based on a population imbalanced DD scheme (Fig. \ref{fig1}) that accumulates effects of inter-state interactions between ultracold atoms in two internal states, without adding any modulation besides the DD itself. We fully characterize the process \cite{PhysRevLett.105.053201}, confirming the prediction of \cite{harber2002effect} that the Bloch sphere is twisted by inter-state interaction. We determine $a_{22}+a_{11}-2a_{12}$ with 0.02 $a_{0}$ uncertainty for both magnetic sensitive and insensitive transitions. Around the equator, this twist generates spin-squeezing. We measured evolution of a state with gaussian noise around the X axis during DD, showing noise squeezing of $2.79 \pm 0.43$ dB with weak non-linear interaction. This is relevant when there are initialization errors, and can be used to increase sensitivity.

Our method can also be used for spin-squeezing beyond the standard quantum limit, similar to \citep{gross2010nonlinear,riedel2010atom}. Long coherence time achievable in our scheme enables us to generate a spin-squeezed state even with weak interactions. It is also relevant for polaron physics \cite{jorgensen2016observationbosepolaron,schirotzek2009observation,kohstall2012metastability,hu2016bosepolaron,PhysRevX.10.041019} and measurements of weak magnetic dipole-dipole interaction \cite{zou2020magnetic}. Imbalanced DD can increase coherence in these measurements, which will, in turn, increase spectroscopic resolution.

 We use a Bose-Einstein condensate (BEC) of ultracold \textsuperscript{87}Rb atoms in states $\ket{1}=\ket{F=1,m_{f_1}}$ and $\ket{2}=\ket{F=2,m_{f_2}}$, where $F$ is the total spin and $m_{f_{1,2}}$ is spin projection along the magnetic field axis. The mean-field energy shifts are:
\begin{equation}
\begin{split}
\delta E_1 =\frac{4\pi\hbar^2}{m} (\alpha_{11}a_{11}n_1+\alpha_{12}a_{12}n_2),\\
\delta E_2 =\frac{4\pi\hbar^2}{m} (\alpha_{22}a_{22}n_2+\alpha_{12}a_{12}n_1),
\end{split}
\end{equation}

$n_{1,2}$ are densities in different states, $a_{ij}$ are \textit{s}-wave scattering lengths, $\alpha_{ij}$  are correlation factors accounting for Bose statistics (for thermal clouds $\alpha_{ij}=2$, and for BEC $\alpha_{ij}=1$ \citep{harber2002effect}), $m$ is atomic mass, and $\hbar$ is the reduced Planck constant. The energy difference in a BEC is,
\begin{equation}
\begin{split}
\delta E_2-\delta E_1 =\frac{2\pi\hbar^2}{m} \biggl((a_{22}-a_{11})(n_1+n_2) \\
+(2a_{12}-a_{22}-a_{11})(n_1-n_2)\biggr).
\end{split}
\label{eq:energy_diff}
\end{equation}
Here, the first term is proportional to the total density $n=n_1+n_2$, while the second is proportional to the density difference $\delta n=n_1-n_2$. 

In a typical Ramsey experiment atoms are in an equal superposition $\frac{1}{\sqrt{2}}\left(\ket{1}+\ket{2}\right)$ during the interrogation time. Therefore, the second term in Eq. (\ref{eq:energy_diff}) is eliminated, leaving only contribution proportional to $n$ in the phase shift. We can prepare a density difference by applying a pulse with area $\theta \ne \pi/2$, generating the state $\cos\frac{\theta}{2} \ket{1}+\sin\frac{\theta}{2}\ket{2}$. Here the density difference is $\delta n=n\cos{\theta}$. Following a hold time $T$ a second pulse with area $\pi-\theta$  converts phase difference $\left(\delta E_2-\delta E_1\right)T/\hbar $ into population difference.

In our imbalanced DD scheme (Fig. \ref{fig1} (a)) we apply a number of echo pulses $N_e$ between two pulses of $\theta$ and $\pi-\theta$. Our pulses have a Rabi frequency $\Omega_R$ and alternating phases of $0^{\circ},\ 90^{\circ}$, rotating the state around the  Bloch sphere's X,Y axes sequentially. This scheme has several benefits. First, it cancels the first term in Eq. (\ref{eq:energy_diff}), because $n$ does not change following a $\pi$ pulse. Second, quasi-static external noises and inhomogeneous dephasing mechanisms (e.g. due to inhomogeneous density or light shift) as well as spatial phase evolution \cite{PhysRevA.80.023603} are nullified, allowing for longer coherence time. The echo pulses reverse the sign of $\delta n$. Therefore, we accumulate phase due to the second term in Eq. (\ref{eq:energy_diff}) and achieve an increased sensitivity to $a_{11}+a_{22}-2a_{12}$, which is typically smaller than $a_{22}-a_{11}$.

The calculated population $P=\frac{N_1}{N_1+N2}$ in state $\ket{2}$ at the end of this sequence, is given by \footnote{See supplemental material}
\begin{equation}\label{P_twist}
 P=
 \frac{1}{4} \biggl[ 2\sin^{2}\theta\cos \left(\phi+ \frac{ g n}{\hbar} N_{e}T\cos \theta \right)+\cos2\theta+3 \biggr]
\end{equation}
where $\phi$ is the last pulse phase and $ g =\frac{4\pi \hbar^2}{m} \left( a_{11}+a_{22}-2a_{12} \right)$ is the interaction shift. This interaction twists the Bloch sphere (Fig. \ref{fig1}(b)), where the upper hemisphere ($\theta<\pi/2$)  and lower hemisphere ($\theta>\pi/2$) rotate in opposite directions. Measured phase shifts (circles) at different latitudes show progress for increasing evolution times  (blue - 12.1 ms, green - 26.2 ms, orange - 54.6 ms).

Interactions during echoes are negligible compared to Rabi frequency ($\frac{gn}{\hbar}\ll \Omega_R$, \cite{Note1}). Rotating the Bloch vector around X,Y cancels noises in control pulses \cite{gullion1990new,de2010universal,wang2012comparison} and also renders the rotations symmetric with respect to the Bloch sphere equator in every XYXY block. Any residual phase accumulated due to population imbalance during the pulses is equal in the upper and lower Bloch hemispheres.

\begin{figure}[t]
%    \centering
    \includegraphics[width=0.95 \linewidth]{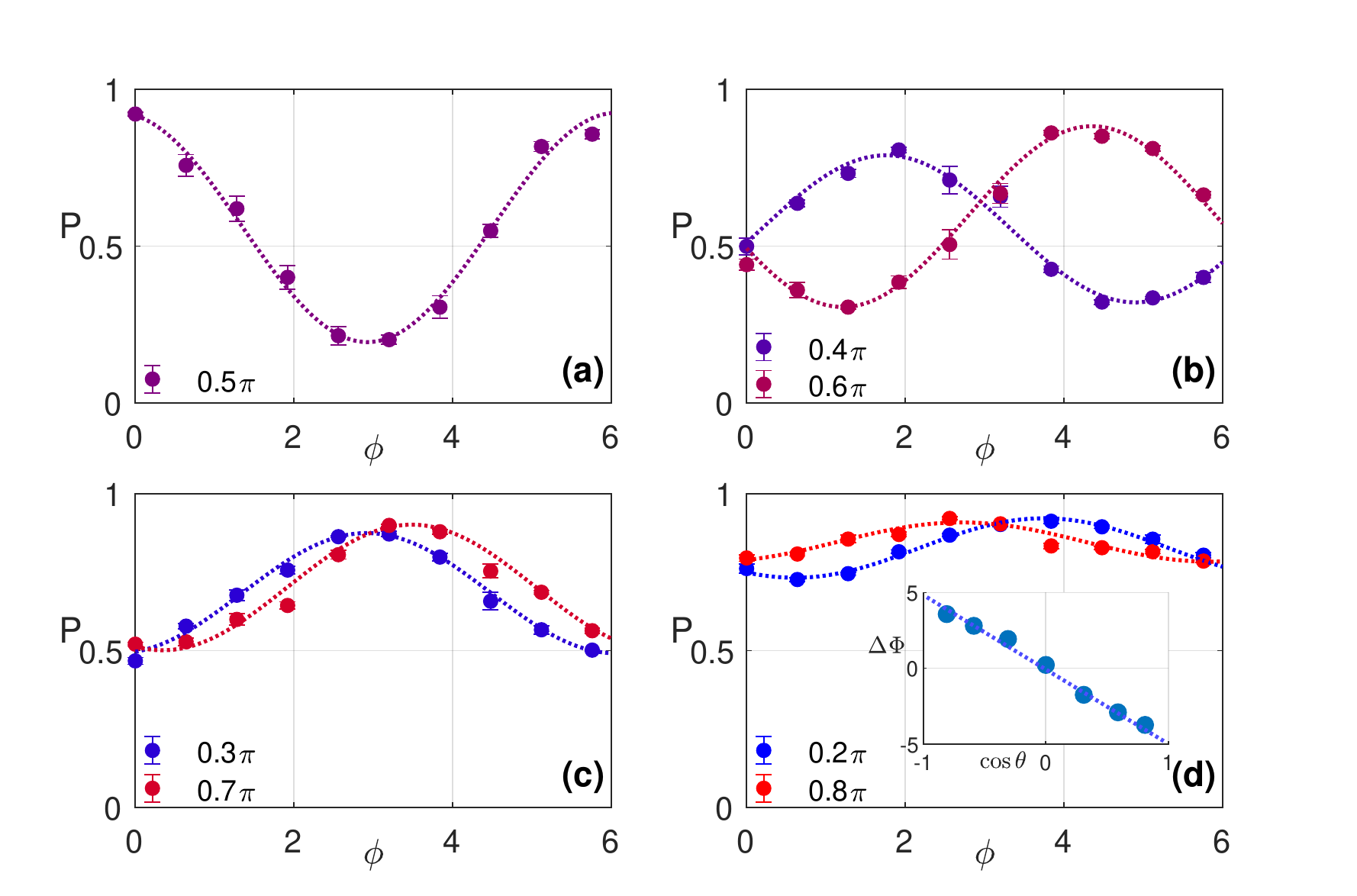}
    \caption{\label{fig2} Imbalanced DD on $\ket{1,0}\leftrightarrow\ket{2,0}$ transition, with 16 echo pulses and hold time $T=2.88$ ms, giving a total time of 92 ms. (a-d) Measured population P (circles) while scanning $\phi$ for different $\theta$, showing clear fringes and high contrast. The fringes are phase shifted due to inter-state interactions and density difference $\delta n/n=\cos{\theta}$. Dotted lines are fits to the data, error bars indicate one standard deviation from four repeated measurements. (d, inset) Phase shift extracted from (a-d) fits (circles) is proportional to $\cos{\theta}$ as predicted by Eq. (\ref{P_twist}), dotted line is a linear fit.}
\end{figure}

In our system we trap a BEC of $\sim 5\times10^5$ \textsuperscript{87}Rb atoms in a crossed dipole trap with trapping frequencies of $\left(\omega_x ,\omega_y, \omega_z\right) = 2\pi\times\left(31,37,109\right)$ Hz. We use MW radiation close to 6.834 GHz to drive transitions between hyperfine levels in a magnetic field of $2.07$ G. We drive both magnetic insensitive transition $\ket{1,0}\leftrightarrow\ket{2,0}$ and magnetic sensitive transitions $\ket{1,1}\leftrightarrow\ket{2,0}$, $\ket{1,1}\leftrightarrow\ket{2,2}$, which are separated by a large Zeeman splitting $\Delta \gg \Omega_R$.

To perform an imbalanced DD scheme, we start with all atoms in state $\ket{1}=\ket{1,0}$ (magnetic insensitive transition) or $\ket{1}=\ket{1,1}$  (magnetic sensitive transitions) and apply MW pulses to couple to state $\ket{2}=\ket{2,0}$ or $\ket{2}=\ket{2,2}$. We apply a pulse of length $t_p$ to rotate the Bloch vector around the X axis with a polar angle $\theta=\Omega_R t_p$. After hold time $T$ we apply a $\pi$ pulse to invert populations. We then alternate between $\pi$ pulses around the X and Y axes with hold time of $2T$ in between (XY8 sequence \cite{ahmed2013robustness}). Finally, after another hold time $T$ we apply a $\pi-\theta$ pulse with phase $\phi$, rotating the vector around the axis $\cos{\phi}\cdot\hat{x}-\sin{\phi}\cdot\hat{y}$. An illustration of our pulse sequence is shown in Fig. \ref{fig1} (a). The train of echo pulses can be prolonged to accumulate more phase and increase sensitivity to $g$. We used up to 72 echo pulses.

\begin{figure}[t]
\includegraphics[width=0.95 \linewidth]
{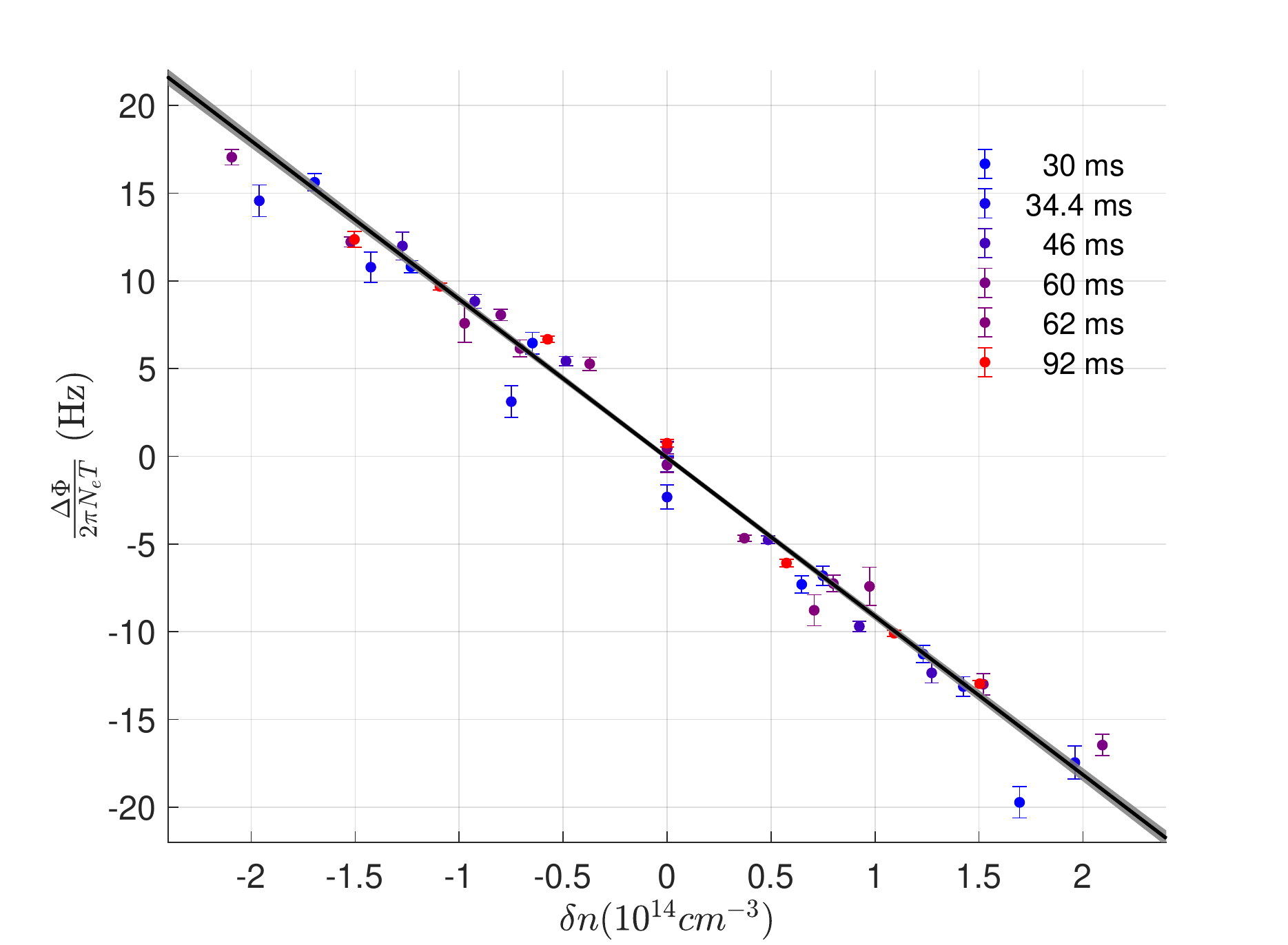}
\caption{\label{fig3} Frequency shift due to interactions between $\ket{1,0} \leftrightarrow \ket{2,0}$ - Measured frequency shift $\frac{\Delta\Phi}{2\pi N_eT}$ and density difference $\delta n$ from measurements of imbalanced DD with various total evolution times $T_{tot}=2N_eT$ (indicated in figure legend). A linear fit (black line, one standard deviation in gray) yields scattering length difference of $\left(a_{11}+a_{22}-2a_{12}\right)_{1,0\to2,0}=-1.17\pm0.02\ a_{0}$, showing the high precision achieved with this method.}
\end{figure}

After the final pulse we release the cloud from the trap and let it expand for 20 ms. We measure population in state $\ket{2}$ using a normalized detection scheme \cite{Note1}. We measure the cloud size and extract chemical potential and average density.

Our measurements on magnetic insensitive transition ($\ket{1,0}\leftrightarrow\ket{2,0}$) after 16 echo pulses with hold time $T=2.88$ ms (total time of 92 ms) are shown in Fig. \ref{fig2} (a-d), filled circles are measured populations vs $\phi$, where different colors indicate different polar angles $\theta$. The dotted lines are fits to $C\cos\left(\phi+\Delta\Phi\right)+B$, where $C$ is the contrast, $\phi$ is the final pulse phase, $\Delta\Phi$ is the phase shift, and $B$ is the bias. The high contrast of  Fig. \ref{fig2} (a) ($\theta=\pi/2$) indicates high coherence after a long DD time of 92 ms, compared with our typical Ramsey coherence time of $\sim30$ ms \cite{PhysRevLett.124.163401}. The contrast decreases with $\theta$, as expected from Eq. (\ref{P_twist}), but the fringes are clear (Fig. \ref{fig2} (b-d)) with excellent agreement with our fits. All measurements shows a phase shift  proportional to $\cos{\theta}$ (Fig. \ref{fig2} (d), inset) as predicted by Eq. (\ref{P_twist}).

\begin{figure}[t]
\center{\includegraphics[width=0.95 \linewidth]
{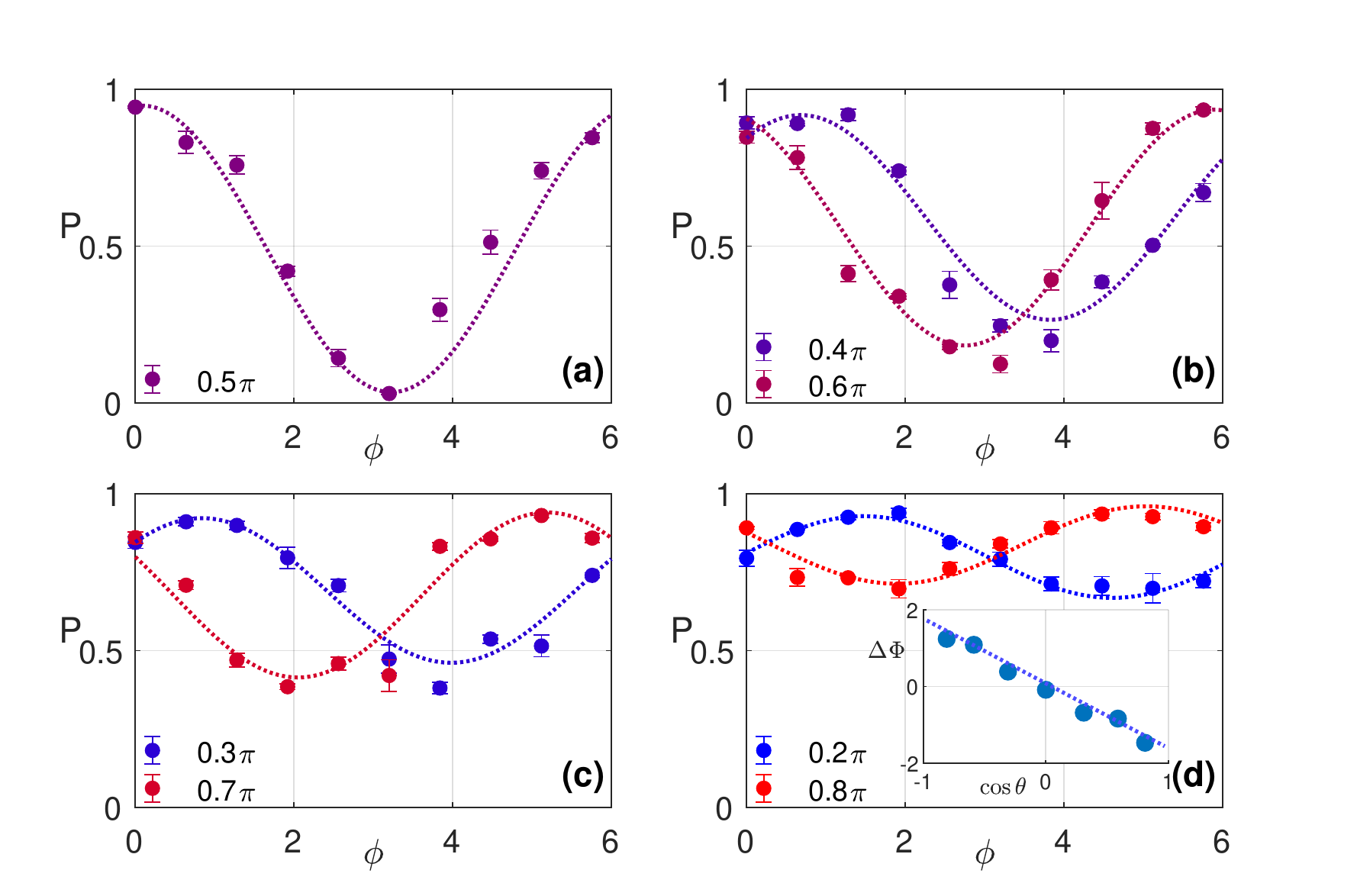}}
\caption{\label{fig4} Imbalanced DD on magnetic sensitive transition $\ket{1,1}\leftrightarrow\ket{2,0}$, with 56 echo pulses and hold time $T=0.38$ ms, giving a total time of $42.4$ ms. Short hold time $T$ is needed to decouple from strong magnetic noises. (a-d) Measured population $P$ while scanning $\phi$ for different $\theta$. Showing fringes with high coherence, even in the presence of strong magnetic noises. (d,inset) Phase shift taken form (a-d) fits (circles),  showing a linear dependence (dotted line) on population imbalance $\cos{\theta}$, similar to the magnetic insensitive measurements (Fig. \ref{fig2}).
}\end{figure}

To validate our measurement and test the performance of this sequence, we repeated this measurement for different times $T_{tot}=2N_{e}T$, varying $T$ and $N_{e}$. Our measurement time was limited by inelastic collisions of atoms in state $\ket{2,0}$, transferring them to states $\ket{2,-1}$ and $\ket{2,1}$ \citep{widera2006precision}.
The frequency shift we measured $\Delta f=\frac{\Delta\Phi}{2\pi T_{tot}}$ grows linearly with $\delta n$ (Fig. \ref{fig3}). From a linear fit we extract $\left(a_{11}+a_{22}-2a_{12}\right)_{1,0\to2,0}=-1.17\pm0.02\ a_{0}$ (statistical error of one standard deviation), showing the high precision and SNR achieved with imbalanced DD. Our main source of systematic errors is density uncertainty, which we estimate to be $ 16 \%$ \cite{Note1,PhysRevLett.124.163401}. Using Ramsey spectroscopy we measured $\left(a_{22}-a_{11}\right)_{1,0\to2,0}=-3.4\pm 0.2 \ a_{0}$, combining the two measurements we get $\left(\frac{a_{11}+a_{22}-2a_{12}}{a_{22}-a_{11}}\right)_{1,0\to2,0}=0.34\pm 0.02$ which is insensitive to density, and comparable with another recent measurement \cite{zou2020magnetic}. This measurement of scattering length difference is useful for calculating interatomic potentials.

\begin{figure}[t]
 \centering
     \begin{overpic}
            [width=0.9 \linewidth]{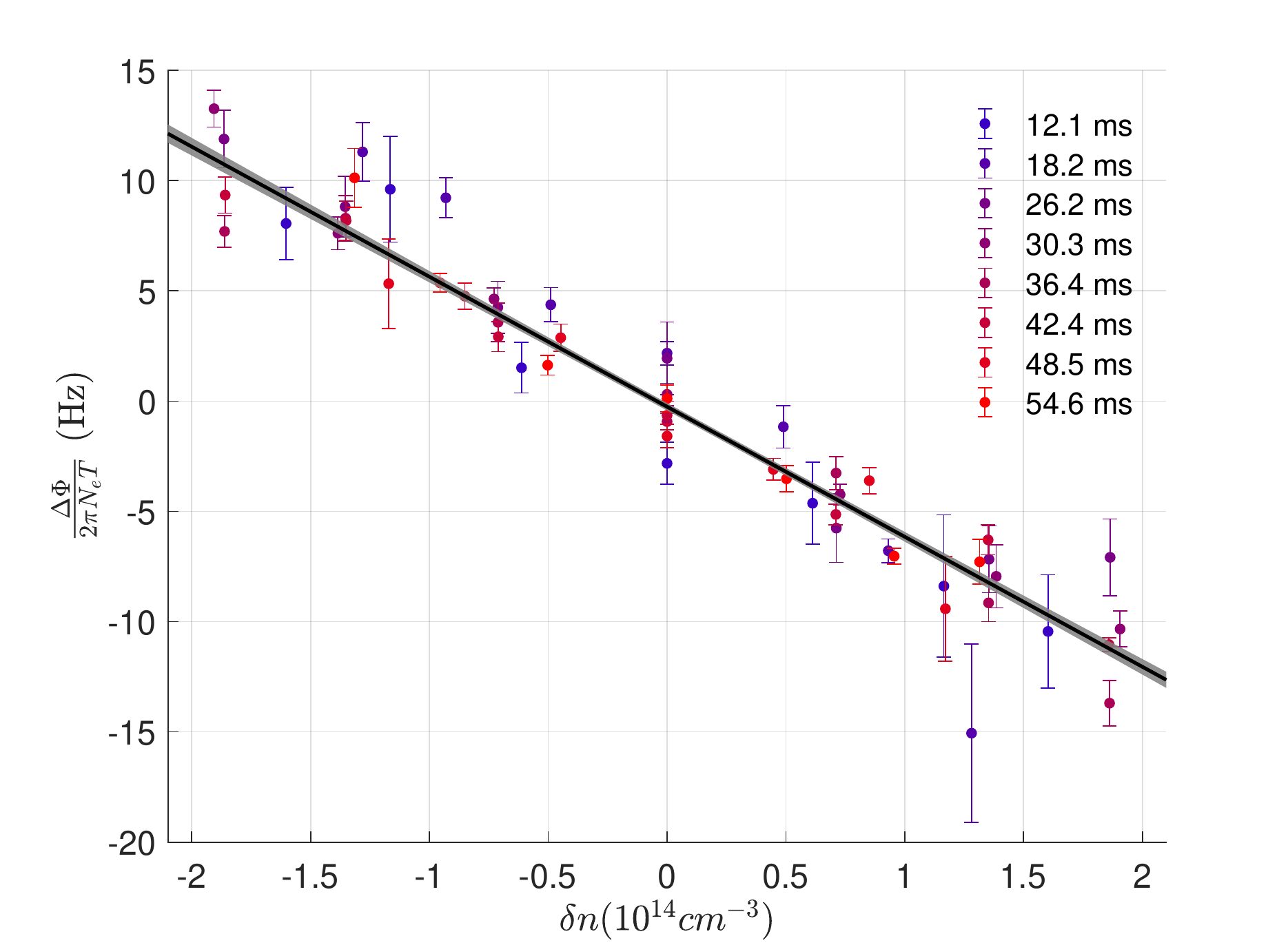}
            \put(30,30){\textbf{(a)}}
        \end{overpic}
        \begin{overpic}
            [width=0.9\linewidth]{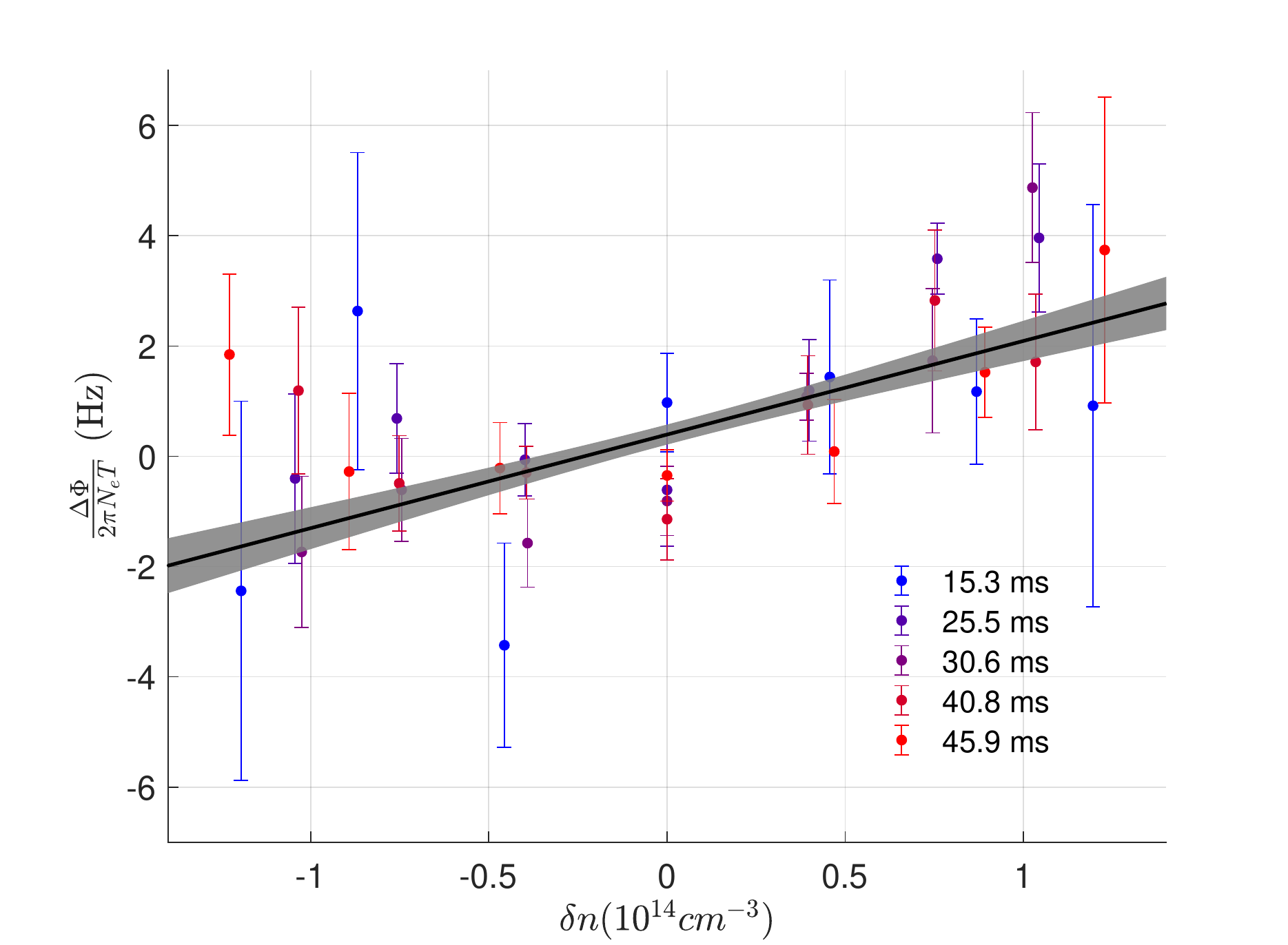}
            \put(30,30){\textbf{(b)}}
        \end{overpic}

\caption{\label{fig5} Frequency shift due to interactions in magnetic sensitive transitions - The measured frequency shift $\frac{\Delta\Phi}{2\pi N_eT}$ and density difference $\delta n$ from imbalanced DD with various $N_e$ and $T$, for $\ket{1,1}\leftrightarrow\ket{2,0}$ (a) and $\ket{1,1}\leftrightarrow\ket{2,2}$ (b)transitions. A linear fit (black line, one standard deviation in gray) gives a scattering length difference of (a) $\left(a_{11}+a_{22}-2a_{12}\right)_{1,1\to2,0}=-0.76\pm0.02\ a_{0}$ and (b) $\left(a_{11}+a_{22}-2a_{12}\right)_{1,1\to2,2}=0.22\pm0.04\ a_{0}$ showing the high precision achieved in this method even in the presence of strong magnetic noise.}
\end{figure}

To show the ability of DD in a noisy environment, we performed the same measurements on magnetic sensitive transitions, $\ket{1,1}\leftrightarrow\ket{2,0}$ and $\ket{1,1}\leftrightarrow\ket{2,2}$ with magnetic field sensitivity of 0.7 kHz/mG and 2.1 kHz/mG respectively. Our main magnetic noise source was the 50 Hz AC from the electrical grid of $1.31 \pm 0.12$ mG peak to peak at the atoms position \cite{Note1}. Our results for the $\ket{1,1}\leftrightarrow\ket{2,0}$ transition with 56 echo pulses and $T=0.38$ ms (Fig. \ref{fig4}), are similar to our results for magnetic insensitive transition. Even in the presence of strong noises our results show long coherence times and phase shifts linear with $\delta n$ (Fig. \ref{fig4}(d), inset). Our Ramsey coherence time for magnetic sensitive transition is limited to a few milliseconds. With DD we were able to measure fringes with high contrast after a total time of $54.6$ ms for $\ket{1,1}\leftrightarrow\ket{2,0}$ transition and $45.9$ ms for $\ket{1,1}\leftrightarrow\ket{2,2}$ transition.

\begin{figure}[t]
\includegraphics[width=0.8 \linewidth]
{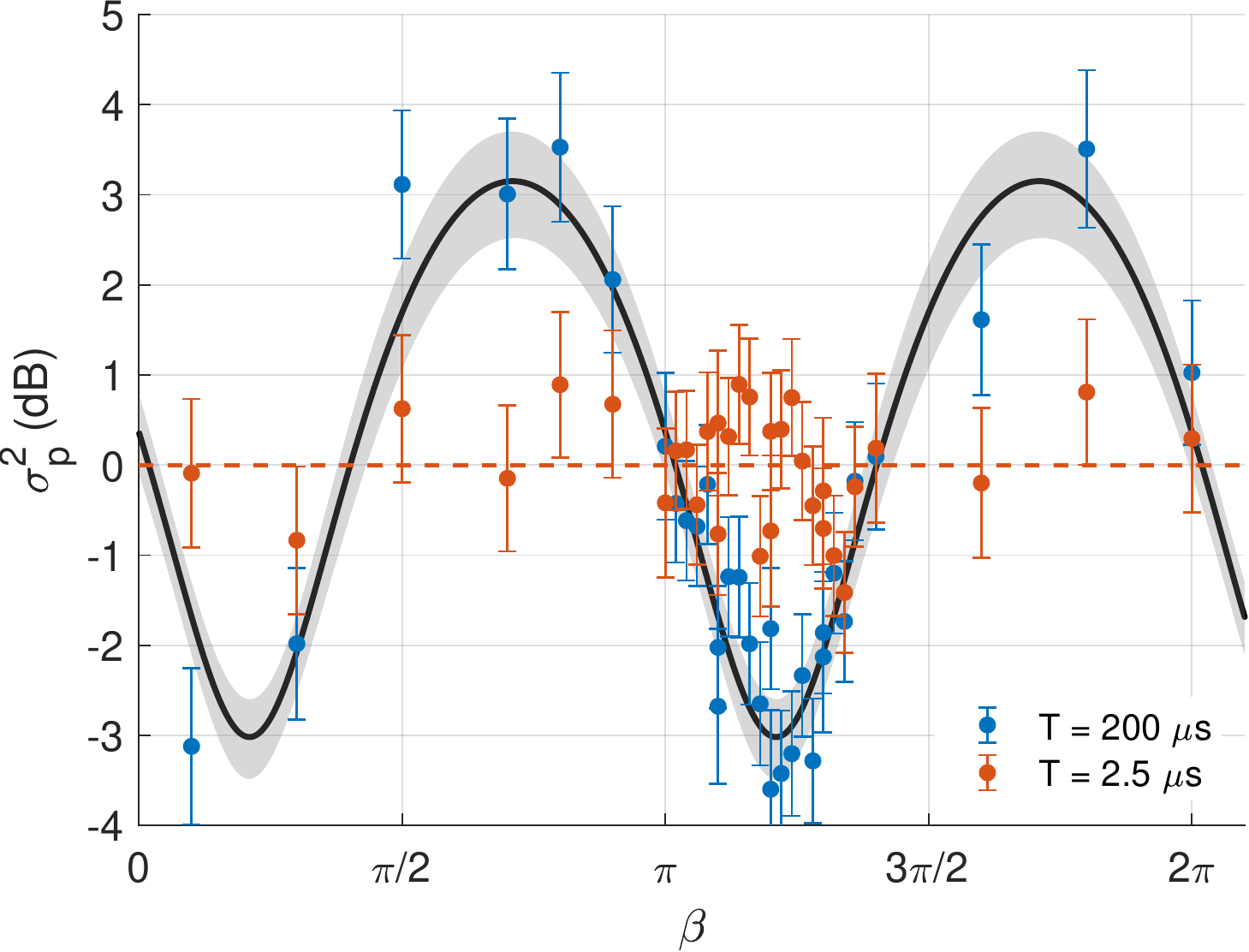}
\caption{\label{fig6} Gaussian noise squeezing. Population variance $\sigma_p^2$ when rotating a state around the X axis with angle $\beta$ after DD with 160 echo pulses. For short hold time ($T = 2.5~\mu$s, red) we get a constant variance, which is used for scaling the vertical axis. For long hold time ($T = 200~\mu$s, blue), we see a double peak feature, using a fit to our data \cite{kitagawa1993squeezed} (black line, one standard deviation in gray) we extract maximal squeezing of $2.79 \pm 0.43$ dB at an angle $\beta_\circ = 218 \pm 14 ^{\circ} $, compared to short hold time mean population variance.}
\end{figure}

We repeated this measurement for different times $T_{tot}$, varying $T$ and $N_{e}$ for the magnetic sensitive transitions, and measured a frequency shift $\Delta f$ that grows linearly with the density difference (Fig. \ref{fig5}). From a linear fit we extract $\left(a_{11}+a_{22}-2a_{21}\right)=-0.76\pm0.02\ a_0$ for $\ket{1,1}\to\ket{2,0}$  (Fig. \ref{fig5} (a)) transition and $\left(a_{11}+a_{22}-2a_{21}\right)=0.22\pm0.04\ a_0$ for $\ket{1,1}\to\ket{2,2}$ transition (Fig. \ref{fig5} (b)), demonstrating high precision and SNR even in a noisy environment. 

When comparing the measurements for all three transitions with calculated scattering lengths \footnote{E. Tiesniga, (private communication)}, our results are within a factor of 2 from the theory for $\ket{1,0}\to\ket{2,0}$ and $\ket{1,1}\to\ket{2,0}$ transitions, and within a factor of 5 for the $\ket{1,1}\to\ket{2,2}$ transition. Our method is applicable close to a Feshbach resonance where typically there are no magnetic insensitive transitions and one of the scattering lengths diverges. This will result in larger phase shifts and can be used to generate spin-squeezed states \cite{gross2010nonlinear,riedel2010atom}.

To show a potential application of our scheme for spin-squeezing we demonstrate squeezing of a state with gaussian noise, which emulates initialization errors. Here we used the magnetic insensitive transition $\ket{1,0}\leftrightarrow\ket{2,0}$. We  prepared a state with a symmetric gaussian noise around the X axis of the Bloch sphere by starting in state $\ket{1,0}$ and applying a pulse of $\frac{\pi}{2} (1+\varepsilon_{\theta})$ with a phase $\frac{\pi}{2} (1+\varepsilon_{\phi})$, where $\varepsilon_{\phi},\  \varepsilon_{\theta}$ are randomly chosen errors ($<1$) from a gaussian distribution, in pulse length and phase, respectively. We carried out a DD scheme of concatenated XY8 blocks with hold time of $2T$ between echo pulses. After this pulse sequence, the state remains centered on the X axis. We then applied a final pulse to rotate the state around the X axis with angle $\beta$ and measured population. We repeated this sequence with different $T$ to see how the twisting dynamics changes the statistical variance in population $\sigma_p^2$ as we scan $\beta$.

 In Fig. \ref{fig6} we show our results for added gaussian noise of 25\% in state preparation and $N_e = 160$. For short times ($T=2.5~\mu$s, $T_{tot} = 0.8$ ms, red) the variance does not change significantly with $\beta$. For long times  ($T=200~\mu$s, $T_{tot} = 64$ ms, blue) we see a double peak feature, with a decrease in variance at angle $\beta_\circ = 218 \pm 14 ^{\circ}$. From a fit to the data we extract $2.79 \pm 0.43$ dB noise squeezing at $\beta_\circ$, comparing short and long times. After decreasing detection noise we have  $3.02 \pm 0.47$ dB noise squeezing (To quantify the detection noise, we measured the population after one $\pi/2$ pulse for different $\beta$). This measurement demonstrates our ability to squeeze symmetric gaussian noise in a specific direction using DD with a weak nonlinear interaction. This method can increase sensitivity in the presence of initialization errors and for spin-squeezing beyond the standard quantum limit.

%, which can be relevant for other systems where DD is a common method, such as NMR experiments or NV centers in diamonds \cite{bar2013solid,de2010universal}. 

In conclusion, we introduced and demonstrated imbalanced dynamic decoupling as a robust method with high contrast and long coherence time. We implemented it to measure interactions in a BEC and the Bloch sphere twist caused by them. Our precision is comparable to other measurements done on a magnetic insensitive transition $\ket{1,-1}\leftrightarrow\ket{2,1}$ \cite{egorov2013measurement,harber2002effect}, and we show similar precision with magnetic sensitive transitions. The scattering length difference we measured for three transitions in \textsuperscript{87}Rb is useful for calculating interatomic potentials. The interactions between imbalanced populations generate a phase shift that is accumulated in our sequence without having to externally perturb the system, as in other DD schemes \cite{shaniv2017quantum,kotler2011single}. This type of interactions are of great interest close to Feshbach resonances, where scattering length diverges, and magnetic field noise is dominant. We demonstrated spin-squeezing of a state with gaussian noise using DD and weak nonlinear interactions, showing reduced uncertainty when rotating our final state in a specific angle. This can be useful for measurements with initialization errors. 

\begin{acknowledgments}
The authors would like to thank Eite Tiesinga, Yotam Shapira and Tom Manovitz for fruitful discussions. This work was supported by the  Israeli Science Foundation, the Israeli Ministry of Science Technology and Space and the Minerva Stiftung
\end{acknowledgments}

%\clearpage

\bibliographystyle{apsrev4-2}
\bibliography{main}

\end{document}

\heading{Supplementary Material for Observation of nonlinear spin dynamics and squeezing in a BEC using dynamic decoupling}
\begin{center} Hagai Edri, Boaz Raz, Gavriel Fleurov,  Roee Ozeri and Nir Davidson\end{center}
\begin{center} Department of Physics of Complex Systems, Weizmann Institute of Science, Rehovot 76100, Israel\end{center}

%\maketitle

\subsection*{Experimental Setup}
In our setup we trap $^{87}$Rb in a crossed dipole trap ($\lambda = $ 1064 nm). We produce an almost pure BEC of $\sim10^5$ atoms with no visible thermal fraction. The trapping frequencies are $\omega_{x,y,z}=2\pi\times \left(31,37,109\right)$.

In each measurement we take two images with imaging light that is resonant with the $F=2 \to F'=3$ transition of the \textsuperscript{87}Rb D2 line after 20~ms time of flight. In the first image we count only the atoms in the state $F=2$. Then, after a short repump pulse ($F=1 \to F'=2$, 40 $\mu$s) that transfers all atoms from $F=1$ to the $F=2$ manifold, we image the cloud again to count the total number of atoms and measure the density. In this way we measure the relative population of state $\ket{2}$ (defined below), without being susceptible to fluctuations in atom numbers between different measurements.

We measure the number density by fitting the integrated column density in the second image with $f(x) = \text{max}\biggl(0,\left( 1-\frac{x^2}{x^2_c}\right)^2\biggr)$, where $x_c$ is the cloud's half width in the strong trapping direction. From which we calculate the chemical potential $\mu$ and the peak number density $n_0=\mu\frac{m}{4\pi\hbar^2a}$,  where $m$ is the atomic mass, $a$ is the scattering length, and $\hbar$ is the reduced Planck constant. The average number density is given by $n=\frac{4}{7}n_0$.

Using this method we are insensitive errors in the calculation of number of atoms and density with Beers law, such as: imaging beam polarization, magnetic field direction, off resonant light and saturation due to high OD. Our method is mostly sensitive to magnification and resolution errors. We calculate the magnification by measuring the atoms fall due to gravity and check our resolution by imaging small clouds in situ, where the cloud is a factor of $\sim 20$ smaller than in our typical image after time of flight.

\subsection*{Dynamic Decoupling}

We use microwave (MW) radiation close to $f_0=6.834$ GHz to induce transitions between hyperfine levels of \textsuperscript{87}Rb atoms, $\ket{1}=\ket{F=1,m_{f_1}}$ and $\ket{2}=\ket{F=2,m_{f_2}}$, where $F$ is the total spin and $m_{f_1,2}$ is its projection on the magnetic field axis. We start with all atoms in state $\ket{1}$ and apply a set of pulses to control their state and accumulate phase due to inter-state interactions. Each pulse of length $t_p$ and Rabi frequency $\Omega_R$ rotates the state vector by an angle $\theta=\Omega_R t_p$ around the axis $\hat{n}=\cos{\phi} \hat{x}-\sin{\phi}\hat{y}$. This rotation can be applied to any state $\ket{\psi}=\alpha \ket{1}+\beta\ket{2}$ with arbitrary $\alpha$ and $\beta$ by an operator $R_{\hat{n}}(\theta)= \exp{(-\frac{i}{2}\theta\Vec{\sigma}\cdot\hat{n})}$. Free evolution of the state for a time $T$ corresponds to a rotation around the $\hat{z}$ axis on the Bloch sphere. The angle of rotation is $\epsilon=\frac{\delta E_{12}}{\hbar}T$, where $\delta E_{12}$ is the energy difference between the two levels (Eq. (2) in the main text).

Our pulse sequence starts with a pulse of $\theta$ around the $\hat{x}$ axis followed by a train of $\pi$ pulses around X-Y-X-Y-Y-X-Y-X axes with a free evolution time 2T between pulses (XY8 sequence), we then apply a pulse of $\pi-\theta$ around an axis $\hat{n}$. Applying all these rotations amounts to:
\begin{equation}
\begin{split}
    M(\theta,\phi)=R_{\hat{n}}(\pi-\theta)R_{\hat{z}}(\epsilon)R_{\hat{x}}(\pi)R_{\hat{z}}(2\Tilde{\epsilon})R_{\hat{y}}(\pi)R_{\hat{z}}(2\epsilon)R_{\hat{x}}(\pi)R_{\hat{z}}(2\Tilde{\epsilon})R_{\hat{y}}(\pi)R_{\hat{z}}(2\epsilon)\\
    R_{\hat{y}}(\pi)R_{\hat{z}}(2\Tilde{\epsilon})R_{\hat{x}}(\pi)R_{\hat{z}}(2\epsilon)R_{\hat{y}}(\pi)R_{\hat{z}}(2\Tilde{\epsilon})R_{\hat{x}}(\pi)R_{\hat{z}}(\epsilon)R_{\hat{x}}(\theta),
\end{split}
\end{equation}
where $\Tilde{\epsilon}$ has inverted populations compared to $\epsilon$, such that $\Tilde{\epsilon}-\epsilon$ is proportional to the difference in densities between states  $\ket{1}$ and  $\ket{2}$, and does not depend on the sum of densities. We calculate the population in state $\ket{2}$ after applying this sequence starting from state $\ket{1}$, 
\begin{equation}
    P = \left|M(\theta,\phi)\ket{1}\right|^2= \frac{1}{4} \biggl[ 2\sin^{2}\theta\cos \left(\phi+ \frac{ g n}{\hbar} N_{e}T\cos \theta \right)+\cos2\theta+3 \biggr]
\end{equation}
where $\phi$ is the phase of the last pulse, $ g =\frac{4\pi \hbar^2}{m} \left( a_{11}+a_{22}-2a_{12} \right)$ is the interaction shift and $n$ is the sum of densities in states $\ket{1}$ and $\ket{2}$. In this sequence, the accumulated phase depends solely on the difference in densities $\delta n = n\cos{\theta}$. 

To verify our calculations we performed measurements with a Ramsey sequence, a single echo pulse and two echo pulses (X,-X rotations) on the $\ket{1,0}\leftrightarrow\ket{2,0}$ transition. Showing good agreement with the calculation and our XY8 measurements.

\subsection*{Evolution during the pulses}

In this calculation we did not take into account any phase accumulated during the pulses. It is a good approximation in our system since $\delta E_{12}/\hbar \ll \Omega_R$ and $t_p \ll T$. To verify it we solved numerically the Bloch equations:
\begin{equation}
   \frac{d\Vec{\sigma}}{dt}= \vec{\Omega} \times\Vec{\sigma} 
\end{equation}
Here $\vec{\Omega}=(\Omega_R,0,\delta)$, and we use $\delta = \frac{g n}{\hbar}\cdot \sigma_z$, a detuning that is proportional to the population difference (we neglect any constant detuning of our oscillator since it will be nullified by the XY8 sequence). For our system parameters $\delta/\Omega_R \sim 5\cdot 10^{-3}$ for the magnetic insensitive transition $\ket{1,0}\leftrightarrow\ket{2,0}$. We compared our numerical solution with our analytic calculation for 16 echo pulses, hold time $T=2.88$ ms, and our setup parameters, when taking into account that there is no phase accumulated during the pulse. Our results show a small angle of $0.05$ rad between the two states on the Bloch sphere at the end of the pulse sequence, after a total rotation of $5.07$ rad.

\subsection*{50 Hz Magnetic noise}

\begin{figure}[tb]
   \centering
    \includegraphics[width=0.75\textwidth]{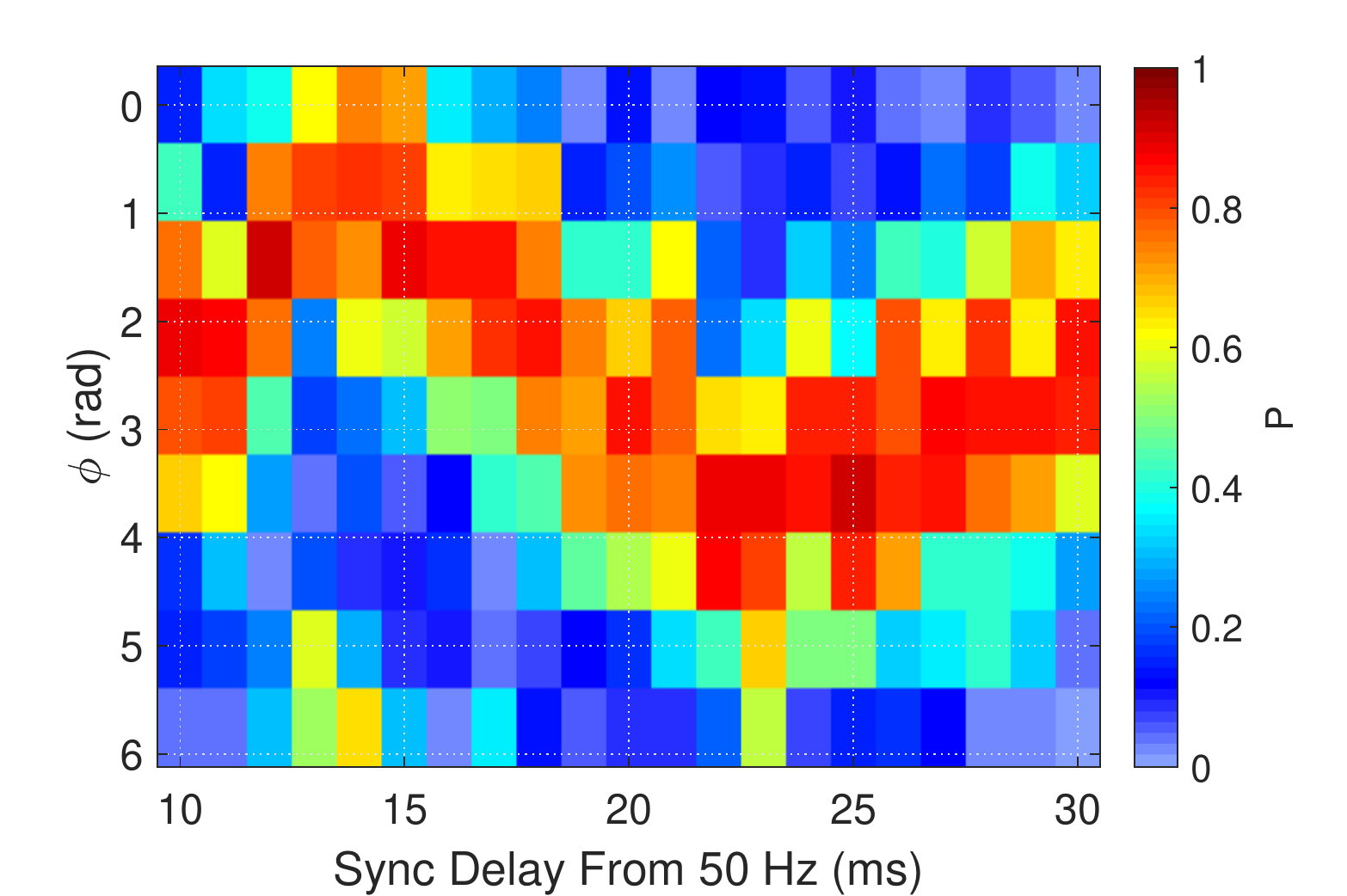}
    \caption{\label{fig1} 50 Hz Noise - Ramsey phase scan with Ramsey time of 0.15 ms, with magnetic sensitive transition $\left|1,1\right\rangle \to\left|2,2\right\rangle$. We changed the delay time between our Ramsey pulses and 50 Hz signal from the electricity grid in our lab. We measured a phase shift that is proportional to the magnetic field noise at 50 Hz as seen by the atoms. The magnetic field at 50 Hz has a $1.31\pm0.12$ mG peak to peak amplitude.}
\end{figure}

An advantage of a dynamic decoupling (DD) scheme is it cancels external noises that are slower than our pulse rate $1/T$, which allows us to measure small frequency shifts in a noisy environment for magnetic sensitive transitions. Our main source of magnetic noise is the 50 Hz signal of the electricity grid in our lab. We measured it using short Ramsey spectroscopy ($0.15$ ms Ramsey time) on the magnetic sensitive transition $\ket{1,1}\leftrightarrow\ket{2,2}$, with our $\frac{\pi}{2}$ MW pulses synchronised to the 50 Hz signal. We changed the delay time between the pulses and the 50 Hz signal and scanned the phase of the second MW pulse. The results are shown in Fig. \ref{fig1}, each column is a Ramsey phase scan fringe. It is clear that the phase of the fringes changes as we change the delay time from the 50 Hz signal (horizontal axis). The 50 Hz noise is clearly seen in our measurement, and it has $1.31\pm0.12$ mG peak to peak amplitude, we do not see any other significant noise frequencies.  To cancel this noise we use DD with short arm time T of $0.35-0.42$ ms when measuring on the magnetic sensitive transition.

\subsection*{Inelastic Collisions}
\begin{figure}[tb]
   \centering
    \includegraphics[width=0.75\textwidth]{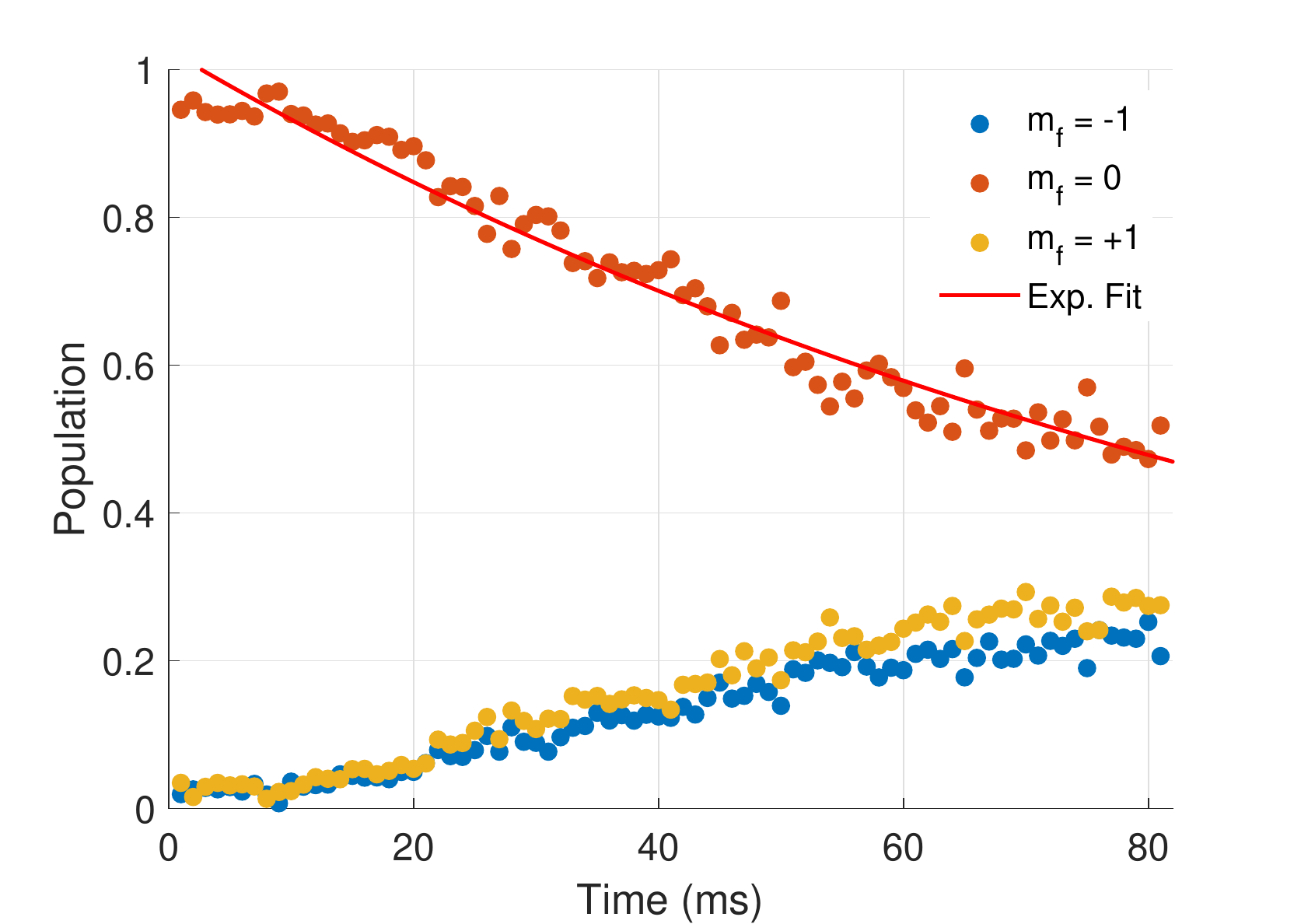}
    \caption{\label{fig2} Inelastic collisions - We measured the population in different spin states $m_f$ of total spin $F=2$. The population in $\ket{2,0}$ decays, while the population of $\ket{2,-1}$ and $\ket{2,1}$ increases almost by the same amount as a result of inelastic collisions of atoms in state $\ket{2,0}$, transferring them to states $\ket{2,-1}$ and $\ket{2,1}$. From an exponential fit we estimate the spin relaxation time to be $105\pm2$ ms. It does not have an effect on the phase shift in our measurement for the magnetic insensitive transition, and only causes a slightly decreased coherence of our fringes.}
\end{figure}

The DD scheme increases our coherence time from $\sim15$ ms to $\sim100$ ms (for magnetic insensitive transition) and from $\sim1$ ms to $\sim50$ ms (for magnetic sensitive transitions). Our main limitation for the magnetic insensitive transition are inelastic collisions of atoms in state $\ket{2,0}$, transferring them to states $\ket{2,-1}$ and $\ket{2,1}$. We can measure the rate of these collisions by holding all atoms in state $\ket{2,0}$ for different times and measuring the  population of different spin states in $F=2$. We do that with a Stern-Gerlach measurement, we release the atoms from the trap and turn on a magnetic field with a constant gradient, after 20 ms time of flight the states are well separated in our images. The result of these measurements are presented in fig. \ref{fig2}. The population in $\ket{2,0}$ decays, while the population of $\ket{2,-1}$ and $\ket{2,1}$ increases almost by the same amount. From an exponential fit we estimate the spin relaxation time to be $105\pm2$ ms. It does not have an effect on the phase shift in our measurement, and only causes a slightly decreased coherence of our fringes, as can be seen in Fig. 2(a) in the main text, where there is a constant bias due to the population of states $\ket{2,-1}$ and $\ket{2,1}$. 

%\subsection*{Magnetic Noise Measurement}

%\subsection*{Coherence Measurement}